\documentclass[8.5pt,twoside,twocolumn]{article}
\usepackage[super,sort&compress,comma]{natbib} 
\usepackage{mhchem}
\usepackage{times,mathptmx}
\usepackage{sectsty}
\usepackage{balance} 

\usepackage{graphicx} %eps figures can be used instead
\usepackage{lastpage}
\usepackage[format=plain,justification=raggedright,singlelinecheck=false,font=small,labelfont=bf,labelsep=space]{caption} 
\usepackage{fancyhdr}
\usepackage{amsmath,amssymb}
\usepackage{subfig}
\usepackage{url}
  
\DeclareMathOperator{\Div}{div}

\begin{document}

\thispagestyle{plain}
\fancypagestyle{plain}{
\fancyhead[L]{\includegraphics[height=8pt]{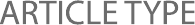}}
\fancyhead[C]{\hspace{-1cm}\includegraphics[height=20pt]{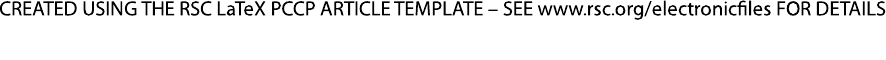}}
\fancyhead[R]{\includegraphics[height=10pt]{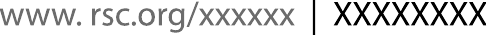}\vspace{-0.2cm}}
\renewcommand{\headrulewidth}{1pt}}
\renewcommand{\thefootnote}{\fnsymbol{footnote}}
\renewcommand\footnoterule{\vspace*{1pt}% 
\hrule width 3.4in height 0.4pt \vspace*{5pt}} 
\setcounter{secnumdepth}{5}

\makeatletter 
\def\subsubsection{\@startsection{subsubsection}{3}{10pt}{-1.25ex plus -1ex minus -.1ex}{0ex plus 0ex}{\normalsize\bf}} 
\def\paragraph{\@startsection{paragraph}{4}{10pt}{-1.25ex plus -1ex minus -.1ex}{0ex plus 0ex}{\normalsize\textit}} 
\renewcommand\@biblabel[1]{#1}            
\renewcommand\@makefntext[1]% 
{\noindent\makebox[0pt][r]{\@thefnmark\,}#1}
\makeatother 
\renewcommand{\figurename}{\small{Fig.}~}
\sectionfont{\large}
\subsectionfont{\normalsize} 

\fancyfoot{}
\fancyfoot[LO,RE]{\vspace{-7pt}\includegraphics[height=9pt]{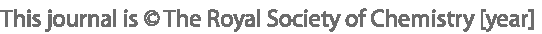}}
\fancyfoot[CO]{\vspace{-7.2pt}\hspace{12.2cm}\includegraphics{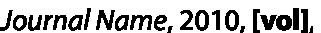}}
\fancyfoot[CE]{\vspace{-7.5pt}\hspace{-13.5cm}\includegraphics{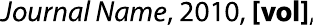}}
\fancyfoot[RO]{\footnotesize{\sffamily{1--\pageref{LastPage} ~\textbar  \hspace{2pt}\thepage}}}
\fancyfoot[LE]{\footnotesize{\sffamily{\thepage~\textbar\hspace{3.45cm} 1--\pageref{LastPage}}}}
\fancyhead{}
\renewcommand{\headrulewidth}{1pt} 
\renewcommand{\footrulewidth}{1pt}
\setlength{\arrayrulewidth}{1pt}
\setlength{\columnsep}{6.5mm}
\setlength\bibsep{1pt}

\twocolumn[
  \begin{@twocolumnfalse}
\noindent\LARGE{\textbf{Rectification properties of conically shaped nanopores: consequences of miniaturization}}
\vspace{0.6cm}

\noindent\large{\textbf{J.-F. Pietschmann,$^{\ast}$\textit{$^{a}$} M.-T. Wolfram\textit{$^{b}$}, M. Burger\textit{$^{c}$}, C. Trautmann\textit{$^{d,f}$}, G. Nguyen\textit{$^{e}$}, M. Pevarnik\textit{$^{e}$}, V. Bayer\textit{$^{f}$} and  Z. Siwy\textit{$^{e}$}}}\vspace{0.5cm}
%Please note that \ast indicates the corresponding author(s) but no footnote text is required. 

% \thanks{MB and VB acknowledge financial support from {\em Volkswagen Stiftung} via the grant {\em Multi-scale simulation of ion transport through biological and synthetic channels}. MTW acknowledges financial support of the Austrian Science Foundation FWF via the Hertha Firnberg Project T456-N23. JFP acknowledges support by the DFG via grant PI 1073/1-1, the German academic exchange service (DAAD) via project 56052884 and the Daimler and Benz fundation via a PostDoc stipend. ZS recognizes the support of the National Science Foundation CHE 1306058}

\date{\today}% It is always \today, today,

\noindent\textit{\small{\textbf{Received Xth XXXXXXXXXX 20XX, Accepted Xth XXXXXXXXX 20XX\newline
First published on the web Xth XXXXXXXXXX 200X}}}

\noindent \textbf{\small{DOI: 10.1039/b000000x}}
\vspace{0.6cm}
%Please do not change this text.

\noindent \normalsize{Nanopores attracted a great deal of scientific interest as templates for biological sensors as well as model systems to understand transport phenomena at the nanoscale. The experimental and theoretical analysis of nanopores has been so far focused on understanding the effect of the pore opening diameter on ionic transport. In this article we present
systematic studies on the dependence of ion transport properties on the pore length. Particular attention was given to the effect of ion current rectification exhibited for conically shaped nanopores with homogeneous surface charges.
We found that reducing the length of conically shaped nanopores significantly lowered their ability to rectify ion current. However, rectification properties of short pores can be enhanced by tailoring the surface charge and the shape of the narrow opening. Furthermore we analyze the relationship of the rectification behavior and ion selectivity for different pore lengths. All simulations were performed using MsSimPore, a software package for solving the Poisson-Nernst-Planck (PNP) equations. It is based on a novel finite element solver and allows for simulations up to surface charge densities of -2 e/nm$^2$. MsSimPore is based on $1$D reduction of the PNP model, but allows for a direct treatment of the pore with bulk electrolyte reservoirs, a feature which was previously used in higher dimensional models only. MsSimPore includes these reservoirs in the calculations; a property especially important for short pores, where the ionic concentrations and the electric potential vary strongly inside the pore as 
well as in the regions next to pore entrance.}
\vspace{0.5cm}
 \end{@twocolumnfalse}
  ]

\footnotetext{MB and VB acknowledge financial support from {\em Volkswagen Stiftung} via the grant {\em Multi-scale simulation of ion transport through biological and synthetic channels}. MTW acknowledges financial support of the Austrian Science Foundation FWF via the Hertha Firnberg Project T456-N23. JFP acknowledges support by the DFG via grant PI 1073/1-1, the German academic exchange service (DAAD) via project 56052884 and the Daimler and Benz fundation via a PostDoc stipend. ZS recognizes the support of the National Science Foundation CHE 1306058}

%Please use \dag to cite the ESI in the main text of the article.
%If you article does not have ESI please remove the the \dag symbol from the title and the above footnotetext.

\footnotetext{\textit{$^{a}$~Numerical Analysis and Scientific Computing, TU Darmstadt, Dolivostr. 15, 64293 Darmstadt, Germany, pietschmann@mathematik.tu-darmstadt.de}}
\footnotetext{\textit{$^{b}$~Department of Mathematics, University of Vienna, Nordbergstrasse 15, 1090 Vienna, Austria}}
\footnotetext{\textit{$^{c}$~Institute for Computational and Applied Mathematics, University M\"unster, Einsteinstr. 62, 48149 M\"unster, Germany}}
\footnotetext{\textit{$^{d}$~Materials Science, TU Darmstadt, Petersenstr. 23, 64287 Darmstadt, Germany}}
\footnotetext{\textit{$^{e}$~Department of Physics and Astronomy, University of California, Irvine, California 92697, United States, zsiwy@uci.edu}}
\footnotetext{\textit{$^{f}$~Materials Research, GSI Helmholtz Center, Planckstr. 1, 64291 Darmstadt, Germany}}

%additional addresses can be cited as above using the lower-case letters, c, d, e... If all authors are from the same address, no letter is required

% \footnotetext{\ddag~Additional footnotes to the title and authors can be included \emph{e.g.}\ `Present address:' or `These authors contributed equally to this work' as above using the symbols: \ddag, \textsection, and \P. Please place the appropriate symbol next to the author's name and include a \texttt{\textbackslash footnotetext} entry in the the correct place in the list.}

\section{Introduction}
Nanopores are nanoscale channels in synthetic materials such as silicon nitride, graphene or polymers, e.g. polyethylene terephthalate\cite{Li2001,Storm2003,Apel2001,Healy2007,Kim2006,Schneider2010,Merchant2010,Garaj2010}. They can be made in a variety of lengths, diameters and shapes, only limited by the thickness
and robustness of the membrane material. The transport properties of the pores can be tuned and modified by the surface charge of the material and by chemical modifications of the channel surface, e.g. via attaching large biomolecules. Due to their versatile and robust behavior, nanopores have emerged as promising tools for regulating the transport of charged particles, sensing single molecules or DNA sequencing \cite{Healy2007-2,Howorka2009,Venkatesan2011,Cherf2012,Manrao2012}.

There is also growing interest in nanopores that rectify ion current. Since these systems feature asymmetric current-voltage characteristics reminiscent of properties of semiconductor diodes, they have the potential to be used as components of ionic circuits for switching and re-directing ionic flow. Channels with diode-like behavior were also the basis for creating ionic logic gates \cite{Han2009,Ali2009}.

One of the first reported nanoporous rectifiers consists of a tapered cone glass pipette or polymer nanopore with negative surface charges \cite{Wei1997,Apel2001,Siwy2002}. The current in these systems is carried primarily by positively charged ions. Higher currents were recorded for voltages of the polarity for which cations moved from the narrow opening to the wide entrance of the pore. Rectifying current-voltage curves were explained via voltage-controlled ionic concentrations in the pore, modeled using the Poisson-Nernst-Planck equations. In the forward bias, for which ionic currents are high, concentrations of both positive and negative ions were found significantly higher than the bulk electrolyte concentration; for the opposite voltage polarity, a depletion zone was created in which concentration of both cations and anions reached values below the bulk concentration.\cite{Powell2009}.\\
Experimental and theoretical studies of rectification in conically shaped nanopores have so far focused mostly on the influence of the pore opening diameter, surface charge density of the pore walls, and shape of the narrowest part of the pores \cite{Ramirez2008,Ali2010}. Influence of the pore length has been experimentally studied only in one publication in which rectification properties of $12$ and $25$ $\mu$m long pores were presented \cite{Apel2011}. The longer pores showed a higher asymmetry in their current-voltage curves compared to the characteristics of the $12$-$\mu$m long channels. In another work \cite{Momotenko2011} the lack of dependence of the rectification on pore length was mentioned. 

In this work we analyze in detail how the rectification property of conically shaped nanopores depends on the pore length in the range between $90$ nm and $12$-$\mu$m. This issue is very important in the development of nanofluidic ionic circuits for lab-on-the-chip and sensory systems. In these devices, both diameter and longitudinal dimensions are often desired to be fabricated on the nanoscale. Previous studies of cylindrically shaped nanopores with a diode surface charge pattern indicated that reducing the pore length from a $\mu$m to a few tens of nanometers caused a significant reduction of the diode rectification \cite{Vlassiouk2008,Vlassiouk2007,Karnik2007}. Thus we investigate robustness of the rectification property of conically shaped nanopores, and identify parameters that could be tuned to improve rectification of short pores.
The paper also discusses in detail the issue of ionic selectivity of conically shaped nanopores and its influence on current-voltage asymmetry.

Currently, mathematical modeling of ionic transport through nanopores is mostly done using the continuum approach based on the Poisson-Nernst-Planck (PNP) equations, which were originally developed in the context of solid state semiconductors \cite{Markowich1990}. The first successful PNP simulations for rectifying current-voltage curves of conically shaped nanopores were reported by Cervera et al.\cite{Cervera2005,Cervera2006,Ramirez2008,Cervera2010,Mubarak2011}. 
The simulations were performed using a one-dimensional ($1$D) reduction of the PNP model together with the Donnan equilibrium values for boundary conditions, and the local electroneutrality requirement. This $1$D approach works well for pores with high aspect ratios, where the applied voltage drops primarily on the nanopore. The same model could not be however used for shorter pores (thin membranes) where so-called access-resistance has to be considered\cite{Hall1975}. In other words, the applied voltage drops not only across the membrane but also at the vicinity to the pore openings. The effect of access resistance was also found important in ion exchange membranes, where it was linked with the nonequilibrium diffuse double-layer at the membrane-solution interface\cite{Manzanares1993}.

Solving 2D and $3$D PNP equations to predict the transport behavior of asymmetric nanopores was reported as well\cite{Liu2007,Constantin2007,White2008}. The advantage of the higher-dimensional analysis is the possibility of seeing the radial dependence of the electric potential and ionic concentrations. It also allows for the explicit treatment of the reservoirs with bulk solution in contact with two pore openings. $3$D PNP simulations require however very high computational power especially for pores with high surface charge densities. The mesh used to discretize PNP is often reduced to $0.1$ nm, which greatly increases the computational costs. Considering high surface charge densities is important for polymer and other types of pores for which the surfaces can carry as much as $-1$ e/nm$^2$.

\noindent The main goal of the study is to understand how rectification of conically shaped nanopores depends on the pore length and which parameters, e.g. surface charge density and pore asymmetry (cone opening angle), will influence the ionic transport. Since $1$D models as defined by Cervera et al. \cite{Cervera2005,Ramirez2008,Cervera2010,Mubarak2011} cannot be used for low aspect ratio pores, and the full $3$D PNP model is difficult to solve for systems with high surface charge densities, we developed a new $1$D PNP approach which is easily applicable to pores of a wide range of lengths and surface charge densities. The new algorithm is implemented in a software-package called MsSimPore, which models ion current through cylindrical, conical and cigar shaped nanopores with highly charged pore walls up to $-2$ e/nm$^2$. Extremely high surface charge densities are important e.g. for reported earlier structures covered with gold whose surface charge in KCl solutions indeed approaches $0.32$ C/m$^2$\cite{Powell2011,Sannomiya2010,Nishizawa1995}. The pores are embedded in membranes with thicknesses from $90$ nm to $12$-$\mu$m. MsSimPore is based on a efficient finite element, hybrid discontinuous Galerkin scheme\cite{Egger2008}, 
which is a novel approach for the simulation of ion transport in nanopores. Electrolyte reservoirs are considered explicitly by the use of Dirichlet boundary conditions, so that short pores can be modeled as well. All calculations can be performed on a standard PC and the package is freely available for download\cite{MsSimPore}.\\
The numercial experiments performed by MsSimPore predict a strong dependence of ion current rectification on the pore length. A $6$ nm wide opening conically shaped nanopore rectifies the current when embedded in a $12$-$\mu$m thick membrane. When the pore length was shortened to $93$ nm, the same pore behaves like an ohmic resistor. Rectification of short pores can be regained by manipulating their surface charge density. An optimum surface charge density is found for which the maximum rectification for a given pore geometry could be achieved. This finding was unexpected because ion current rectification was thought to increase with the increase of the pore ion selectivity\cite{Wei1997}. We find the relation between rectification and selectivity to be more complex. 

\section{Methods} 
\subsection{Modeling and simulation}
\begin{figure}[h!]
\begin{center}
\includegraphics[scale=0.48]{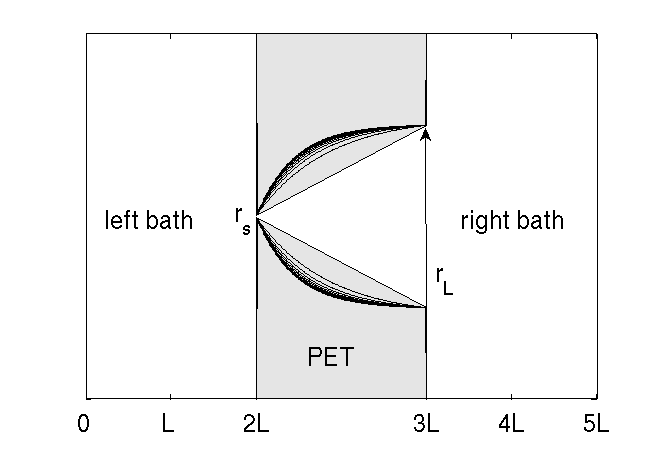}
\caption{Sketch of nanopore separating two electrolyte reservoirs. Possible area functions are shown (not in scale), where the pore shapes correspond to different
values of the parameter (L/h). The curved shape of the small pore opening depends on the ratio of (L/h), the smaller the
ratio the more linear the interpolation between the small and the large opening radius.}\label{f:area}
\end{center}
\end{figure}

The developed approach allows to predict current-voltage curves, together with distributions of ionic concentrations and electric potential along the pore axis in pores with high radial symmetry \cite{Liu2009}. The model in its present form is applicable to cylindrical, conical as well as cigar shaped pores. The main driving forces in the classical PNP equations are diffusion and electrostatic interactions with other ions, as well as surface charges on the pore walls. \\
\noindent In analogy with the experimental setup, MsSimPore simulates a single pore of length $L$ separating two electrolyte solutions. The electrolyte may have different concentration on the left- and right-hand-side. The small and large opening radius of the conical pore is denoted by $r_s$ and $r_l$, respectively. The computational domain considered is $\Omega$ is defined as the interval $[0,5L]$, i.e. the pore separates two electrolyte solutions and each reservoir has a size of $2L$ (this is usually sufficient for equilibration) attached to the right and left, see Fig. \ref{f:area}. The electric potential is given by $V = V(x)$, where $x$ corresponds to the position along the pore axis. The small opening of the pore is positioned at $x=2L$, the large at $x=3L$. The concentration of each ionic species present in the electrolyte is $\rho_i = \rho_i(x)$, $i=1, \ldots m$. The reduced PNP model reads as:
\begin{subequations}\label{e:classpnp}
\begin{align}
&-\Div( \varepsilon A(x) \nabla V) = e A(x)\sum_i z_i \rho_i + \partial A(x) \sigma(x)\\
&0 = \Div(A(x) D_i(x)  (\nabla \rho_i +  z_i \frac{e}{k_B T} \rho_i \nabla V)),
\end{align}
\end{subequations}
where  $A = A(x)$ describes the cross-section of the pore  and $\partial A = \partial A(x)$ its circumference. Here $\varepsilon$ denotes the dielectric coefficient, $e$ the elementary charge, $k_B$ the Boltzmann constant, $T$ the temperature, $z_i$ and $D_i$ the valence and diffusivity of each ionic species, respectively. The function $\sigma = \sigma(x)$ corresponds to the surface charge inside the pore.\\
The area function is defined as follows: The function
\begin{align*}
&r(x)=\\
&\frac{r_l-r_s\exp(-(L/h)^n) - (r_s-r_l) \exp(-((x-2L)/L)^n (L/h)^n) }{1-\exp(-(L/h)^n)},
\end{align*}
interpolates between $r_s$ and $r_l$ inside the pore region $[2L,3L]$. Thus, the area function inside the pore region is given by
\begin{align*}
 A(x) = r(x)r(x)\pi,
\end{align*}
and takes fixed, large values in the bath regions\cite{Ramirez2008}. This definition implies in particular that $A(x) = r_s^2 \pi$ at $x=2L$ and $A(x) = r_l^2 \pi$ at $x=3L$. The ratio $(L/h)$ and the parameter $n$ determine the curved shape of the pore. If $(L/h) \rightarrow 0$, the area function $A = A(x)$ corresponds to the linear interpolation between circles of radius $r_s$ and $r_l$, modeling a conical pore. For large ratios the shape of the pore is more curved, looking like a cigar, see also Fig. \ref{f:area}.\\ 
The bath concentrations of each ionic species are modeled by Dirichlet boundary conditions, hence we have $\rho_i(x) = \tilde{\rho}_i$ at $x = 0$ and $x=5L$ for each $i=1,\ldots m$, where $\tilde{\rho}_i$ denotes the molar concentration of each ionic species in the bath. Also the applied voltage $V_{appl}$ is modeled via a Dirichlet boundary condition, i.e. $V = V_{appl}$ at $x=0$ and $x=5L$.\\
\noindent Two well known reformulations of Eq. \eqref{e:classpnp} can be found in the literature, either based on the Slotboom variables or the entropy variables (also known as quasi Fermi potentials in the semiconductor community). The Slotboom variables, $u_i =\rho_i \exp(c z_i V)$ guarantee positive concentrations, but the exponentials can cause overflow problems for large applied voltages $V_{appl}$. This problem can be avoided using entropy variables $\varphi_i = \log \rho_i + c z_i V$. Then Eq. \eqref{e:classpnp} reads as 
\begin{subequations}\label{e:fermipnp}
\begin{align}
-\lambda^2 \Div( \varepsilon A(x) \nabla V) &= \kappa A(x) \sum_i z_i \rho_i + \partial A(x) \sigma(x)\\
0 &= \Div(D_i(x) A(x)(\rho_i \nabla \varphi_i)).
\end{align}
\end{subequations}
Here  $$\lambda^2 = \frac{\varepsilon \tilde{A}\tilde{V}}{L^2 \tilde{\sigma} \partial \tilde{A}},\quad \kappa = \frac{e \tilde{\rho} \tilde{A}}{\tilde \sigma \partial \tilde{A}}\quad
\text{ and }\quad c = eV/k_BT$$
are scaling parameters, and $\tilde{V}, \tilde{A}, \ldots$ denote typical values of the physical constants, see Table \ref{t:physpar}. This non-dimensionalization allows stable and unit independent simulations.\\
\begin{table}[h!]
\begin{center}
\caption{Parameters for computation}
		\begin{tabular}{lll}
		Meaning & Value & Unit \\\hline
		Boltzmann constant $k_B$ & $1.3806504 \times 10^{-23}$ & $J/K$ \\
		Vacuum permittivity  $\epsilon_0$ & $8.854187817\times10^{-12}$ & $F/m$ \\
		Relative permittivity $\epsilon_r$ & $78.4$ & \\
		Elementary charge $e$ & $1.602176\times 10^{-19}$ & $C$\\
                Temperature $T$ & 293.16 & $K$\\
		Typical length $\tilde{L}$ & $1$ & $nm$ \\
		Typical concentration $\tilde{c}$ & $3.7037\times 10^{25}$ & $N/l$ \\
		Typical voltage $\tilde{V}$ & $100$ & $mV$\\
		\end{tabular}
		\label{t:physpar}
\end {center}
\end{table}
We solve system \eqref{e:fermipnp} on $\Omega = [0,5L]$, where $[0,2L]$ and $[3L,5L]$ correspond to the left and right bath respectively, $[2L,3L]$ is the pore region. Equation \eqref{e:fermipnp} is discretized using a hybrid discontinuous Galerkin method with upwind stabilization\cite{Egger2008}. This stabilization ensures stability of the numerical scheme for large applied voltages. The discrete nonlinear problem is solved by Newton's method. The calculation of the Newton update is based on the 
non-symmetric solver MUMPS\cite{MUMPS1,MUMPS2}. \\
\noindent MsSimPore has been implemented within the finite element framework of Netgen/NgSolve\cite{Schoberl1997}. It allows the simulation of conical and cigar shaped nanopores for up to six ionic species present in the bath. The graphical user interface distinguishes between pore related input parameters (e.g. small and large opening radius $r_s$ and $r_l$, surface charge $\sigma$,...) and the experimental conditions (e.g. number of species present in the bath and their respective concentrations $\tilde{\rho}_i$, applied voltage $V_{appl}$, temperature....).  We assume that the surface charge $\sigma$ is constant inside the pore, which seems a reasonable assumption for the pores considered. The ion specific parameters, like the valence and the diffusion coefficient, are stored in a small data base, which can be modified and extended by the user. MsSimPore offers several solver options such as simulations for one particularly chosen applied voltage as well as the calculation of current-voltage (IV) and 
rectification curves. The solver output variables are displayed in the graphical user interface and stored in a neutral format for subsequent personal use.\\
MsSimPore uses an adaptive mesh, i.e. a coarse discretization of the computational domain $\Omega$ in the bath regions (using a mesh size of $h=50$ nm), which we refine (as small as $h_{min} = 0.1$ nm) around the narrow tip to resolve the fine features correctly. This automatic refinement reduces the computational costs and allows faster simulations. 
% MsSimPore is available for download at the University of M\"unster\cite{MsSimPore}.

\subsection{Experimental}\label{e:experiments}
The nanopores were fabricated by irradiating $12$-$\mu$m thick polyethylene terephthalate (PET) foils with exactly one single heavy ion. The kinetic energy of the ions was in the GeV range which is sufficiently large to penetrate through the entire PET foil. Along its trajectory, each projectile produces a so-called ion track consisting of damaged material of few nm in diameter\cite{Adla2003}. The track in the foil is converted into an open channel by chemical etching\cite{Sertova2009,Cornelius2007}. For this, the irradiated foil is mounted between two chambers of a custom-made conductivity cell, with one chamber being filled with $9$ M NaOH and the other one filled with a neutralizing solution\cite{Siwy2003}. Given by the high NaOH concentration, dissolving the track from one side, conical nanochannels are created. Replacing the etchant by an electrolyte, current-voltage curves were recorded in the same cell using Ag/AgCl electrodes (chloridated Ag wires), a Keithley 6487 picoammeter/voltage source, and 
various KCl solutions (stock solution of $1$ M KCl, lower concentrations prepared by dilution). The electrode at the small opening 
of 
the pore was grounded, while the other electrode, placed in the cell chamber with the large opening of the pore, was used to apply a given transmembrane potential with respect to the ground electrode. As a result of the heavy ion irradiation and chemical etching, carboxyl groups are created at the pore walls at an estimated density of $\approx -1$ e per nm$^2$, cf. \cite{Wolf2002}.

\section{Results and discussion} 

\subsection{Influence of the surface charge on the rectification behavior}

\begin{figure}[h!]
\centering
% \begin{center}
\includegraphics[width=0.5 \textwidth]{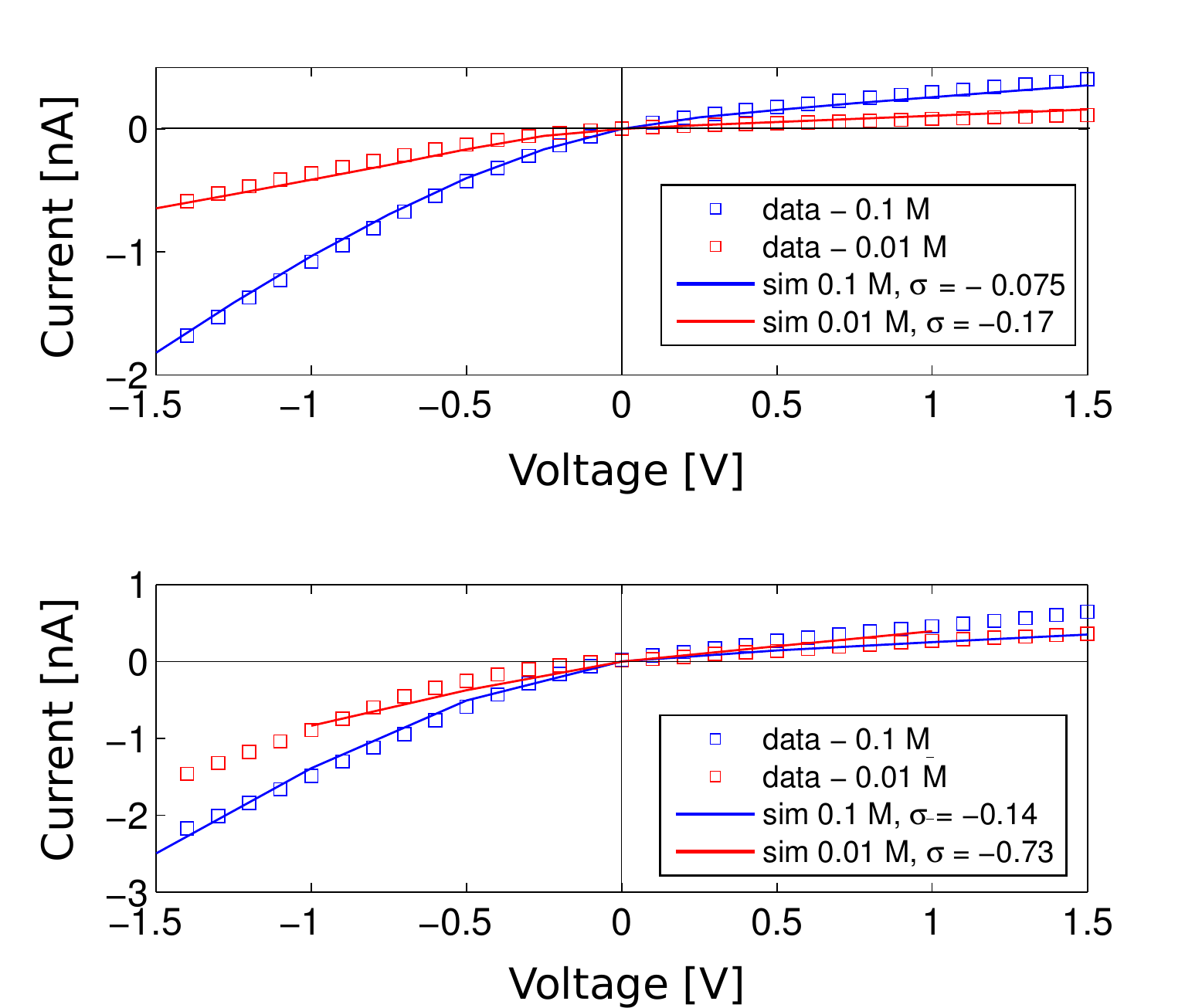}
\caption{Simulated (solid line) and experimental (points) IV curves for a conical nanopore with diameters $d_s = 6$ nm and $d_l = 812$ nm. The data were recorded at symmetric KCl concentrations of $0.1$ and $0.01$ at pH $8$ (upper panel), and pH $5.5$ (lower panel). Numerically found values of surface charge density $\sigma$ of the pore walls are given in the legends as a fraction of the elementary charge.} \label{f:6nm812nm}
% \end{center}
\end{figure}

We start with an investigation of the influence of the surface charge density of pore walls on the rectification behavior of conically shaped nanopores. Fig. \ref{f:6nm812nm} presents experimental data of current-voltage curves through a single conically shaped nanopore with opening diameters of $6$ nm and $812$ nm and a length of $12$-$\mu$m. 
The small pore opening diameter is determined by relating the pore resistance, given by a linear part of an I-V curve in the range between $-100$ mV and $+100$ mV, with the pore geometry. The measurements of the small opening diameter were performed in a high electrolyte concentration of $1$ M KCl in which the surface charges were largely screened. Although the surface carboxyl groups are protonated at pH $3$, recordings in acidic solutions were generally avoided since a few nm in diameter nanopores with neutral walls were occasionally unstable. The big opening diameter is found from a non-specific rate of the material etching which for PET and $9$ M NaOH is $2.13$ nm/min\cite{Apel2001}. Subsequent recordings were performed in two different bulk KCl concentrations of $0.1$ M and $0.01$ M and two pH values of $8$ and $5.5$. The pore walls contain carboxyl groups whose degree of dissociation is higher in more basic conditions. These experimental conditions determine the simulation parameters of the developed $1$D algorithm, treating the surface charge density as the 
parameter to be fitted. The model simulations reproduce the experiments both qualitatively and quantitatively correctly, predicting a decrease of the surface charge density at the solution of pH $5.5$ compared to the surface charge at pH 8 (Fig. \ref{f:6nm812nm}).

Observing rectification behavior in $0.1$ M KCl for a pore with an opening diameter of $6$ nm might seem surprising, but experimental data of rectified ion current for even wider pores were reported\cite{Kovarik2009}. According to the classical Debye-Hueckel theory, the thickness of the electrical double-layer is in this case only $\approx$ $1$ nm, thus the majority of the pore cross-section should be filled with bulk electrolyte. The Debye-Hueckel approximation was however derived for low surface potentials, thus here we perform a detailed numerical analysis of the dependence of the current rectification on pore walls surface charge density between $0$ up to $-2$ e per nm$^2$. We would like to note that modeling a similar set of conditions using $3$D PNP for a full pore length of $12$-$\mu$m is computationally quite expensive.
\begin{figure}[h!]
% \centering
% \begin{center}
\subfloat[\label{f:sel1_6nm812nm}]
% \subfloat[Rectification given by the radio of ionic current at $+1$ V and $-1V$ for a pore with opening diameters of 6 nm and 812 nm in 0.1 M KCl. \label{f:sel1_6nm812nm}]
{\includegraphics[width=0.50 \textwidth]{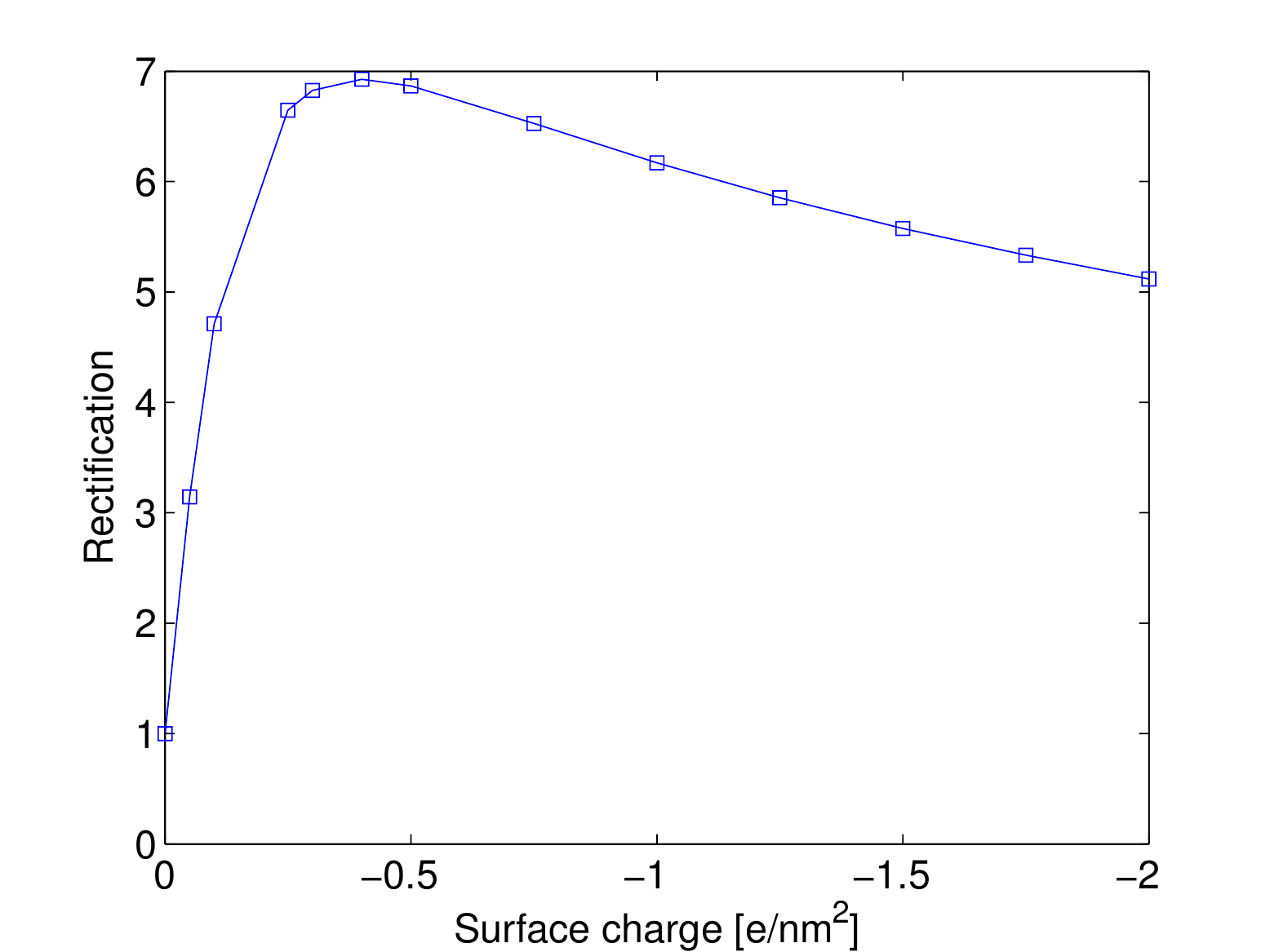}}

\subfloat[\label{f:sel2_6nm812nm}]
% \subfloat[Selectivity of ion currents defined as a ratio of the flux generated by the potassium ions $I^+$ at $0.1$ V of applied transmembrane potential. \label{f:sel2_6nm812nm}
{\includegraphics[width=0.50 \textwidth]{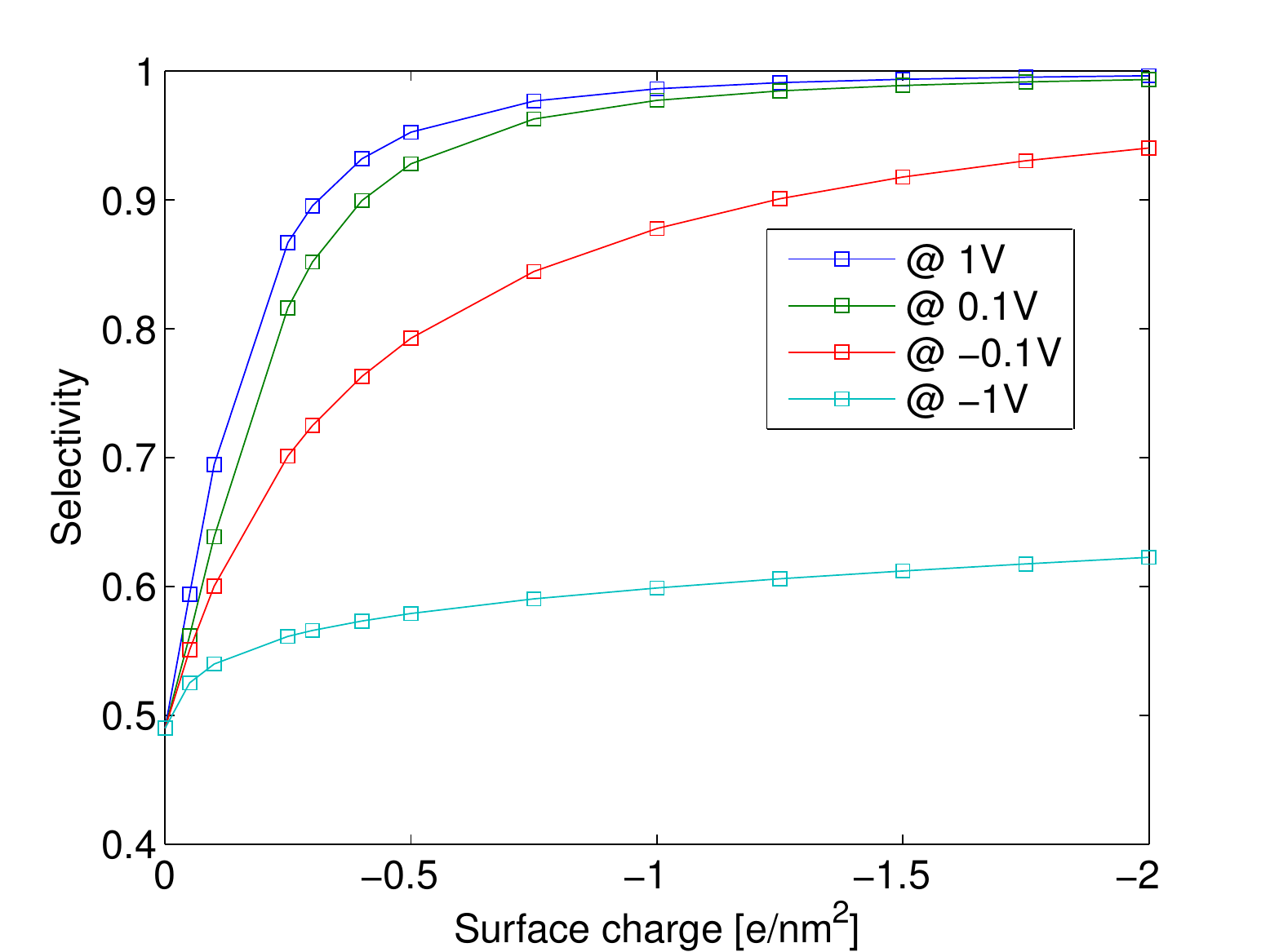}}
% \caption{Rectification and selectivity as a function of the surface charge of a conical nanopore with diameters $d_s = 6$ nm and $d_l = 812$ nm, with symmetric bath concentrations of $0.1$ M KCl. 
\caption{(a) Rectification given by the ratio of ionic currents at $-1$ V and $+1$ V for a pore with opening diameters of $6$ nm and $812$ nm in $0.1$ M KCl; (b) Selectivity of ion currents defined as a ratio of the flux generated by the potassium ions divided by the total current, at different values of applied transmembrane potentials.} \label{f:sel_6nm812nm}
% \end{center}
\end{figure}

\begin{figure}[h!]
\subfloat[\label{f:densities_surfcharge_low}]{\includegraphics[width=0.55\textwidth]{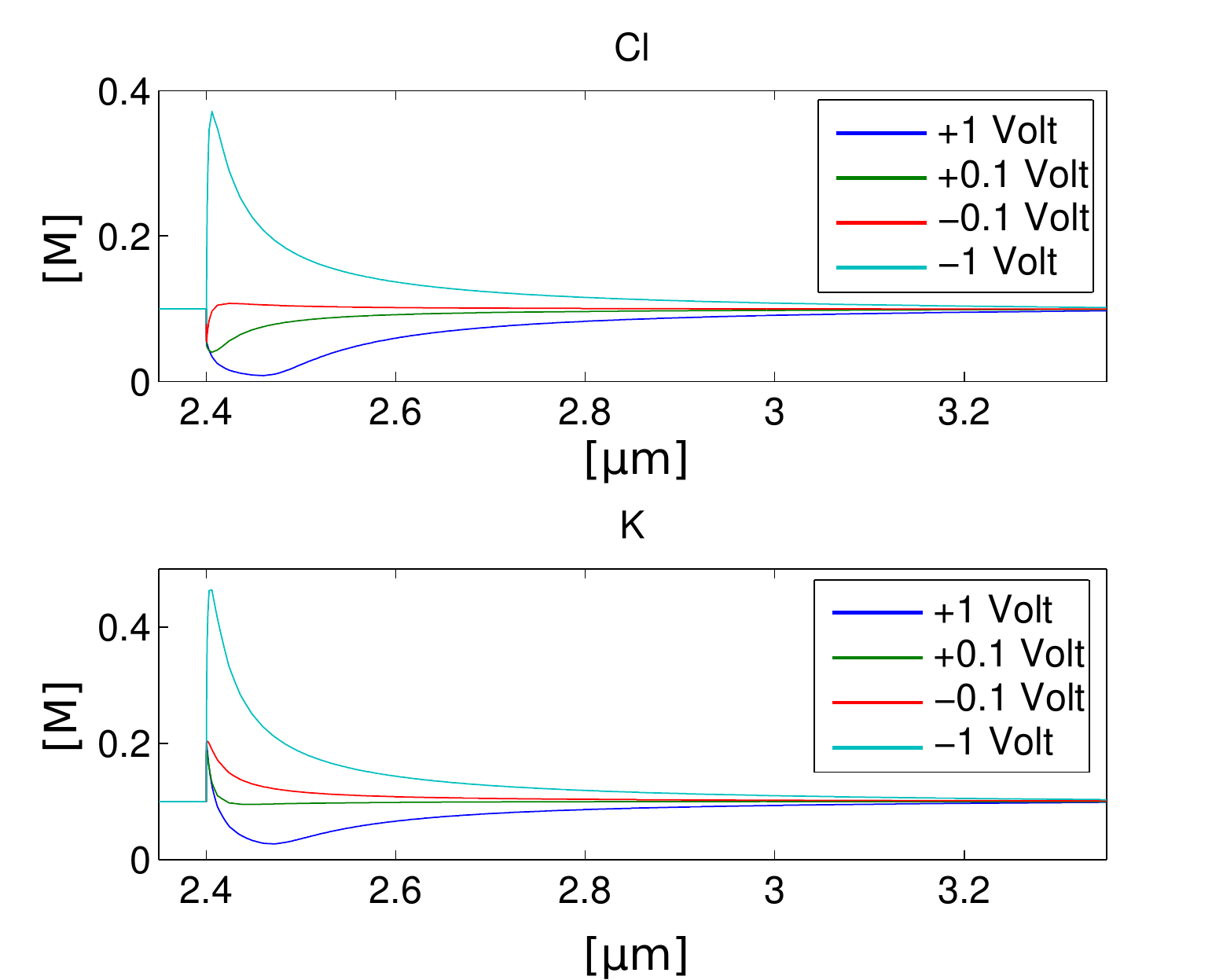}}

\subfloat[\label{f:densities_surfcharge_high}]{\includegraphics[width=0.55\textwidth]{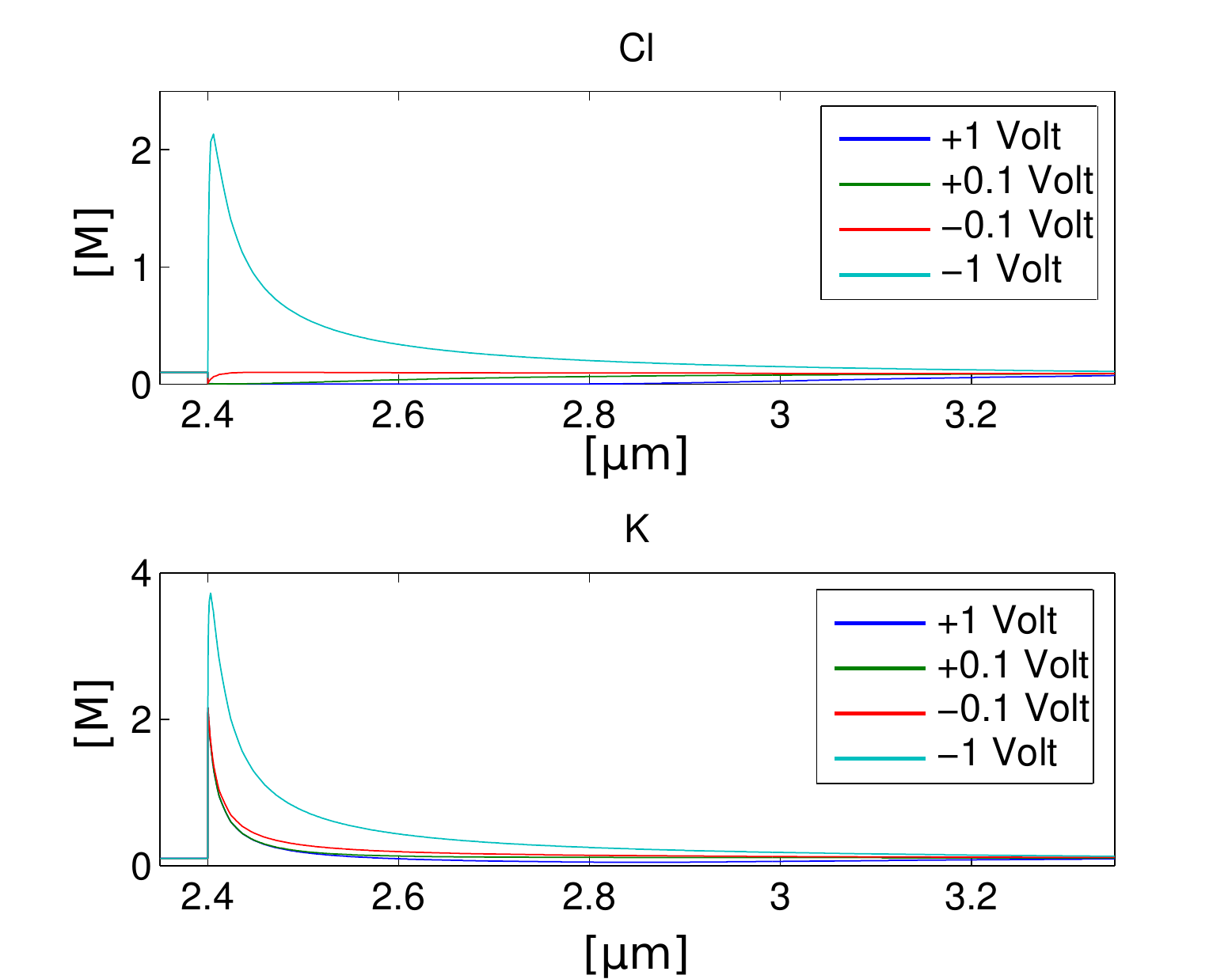}}
\caption{Ionic concentrations along a pore axis for two 12-$\mu$m long conically shaped nanopores (only the pore region is shown) with openings of $6$ nm and $812$ nm and surface charge densities (a) $\sigma = -0.14$ e/nm$^2$ and (b) $\sigma = -2$ e/nm$^2$ and different voltages as indicated in the figure.}
% \caption{Densities inside the pore region for a conical pore with opening diameters $d_s = 6$nm and $d_l = 812$nm for surface charge densities $\sigma = -0.14\;e/nm^2$ and $\sigma = -2\;e/nm^2$ for voltages -1 V, -0.1 V, 0.1 V and 1 V}
\label{f:densities_surfcharge}
\end{figure}

Fig. \ref{f:sel_6nm812nm} shows results of the numerically simulated ion current rectification, calculated as the ratio of currents recorded at -1V and +1V as a function of surface charge density of the pore walls. As expected and shown before experimentally, a conical nanopore with neutral pore walls does not rectify the current\cite{Wei1997}. The modeling revealed a strong dependence of the rectification degree on the surface charge density in the range from 0 up to $\approx$ $-0.4$ e/nm$^2$ where the rectification plateaued. Further increase of the surface charge led to a subsequent decrease of the current-voltage asymmetry, see Fig. \ref{f:sel1_6nm812nm}. \\
In order to explore this observation more, we investigated to which degree each ion, potassium and chloride, contributed to the ion current. In other words we quantified ion selectivity, which is defined as the current carried by potassium ions divided by the total simulated current, Fig. \ref{f:sel2_6nm812nm}. As expected, higher surface charge densities were predicted to make the pore more cation selective, and at low voltages increasing the charge density to $-2$e per nm$^2$ causes a complete exclusion of anions in 6 nm pores. For the data shown in Fig. \ref{f:6nm812nm} the fitted surface charge density of the pore walls was $\approx \,-0.14\,\text{e/nm}^2$ which at positive voltages is indeed sufficient to render the pore almost $70$ \% cation selective.\\
Ion selectivity is however strongly dependent on the voltage magnitude and polarity (Fig. \ref{f:sel2_6nm812nm}). Currents for positive voltages remain cation selective even at $1$ V, while increasing the magnitude of the negative voltages reduces cation selectivity to a value of $0.6$ at $-1$ V even for the highest considered surface charge density. The change of the ion selectivity with voltage can also be clearly seen from profiles of ion concentrations along the pore axis shown in Fig. \ref{f:densities_surfcharge}. For $-1$V, concentration of potassium approaches the concentrations of chloride ions, and concentrations of both ions can exceed the bulk values by more than one order of magnitude.

Comparison of Fig. \ref{f:sel_6nm812nm}(a) and Fig. \ref{f:sel_6nm812nm}(b) suggests that the highest ion current selectivity does not necessarily assure the highest rectification degrees of conically shaped nanopores. In order to understand this effect better, in Fig. \ref{f:densities_surfcharge_low} and Fig. \ref{f:densities_surfcharge_high} we plotted profiles of ionic concentrations along the pore axis for $-0.14$ e/nm$^2$ and $-2$ e/nm$^2$, at different applied voltages of $-1$ V, $+1$ V, $-0.1$ V and $+0.1$ V. Rectification arises from the differences in the ionic distributions for voltages of two opposite polarities. For a pore with surface charge density of $-0.14$ e/nm$^2$, negative voltages caused an increase of ionic concentrations above the bulk values, while for positive voltages there is a zone with depleted concentration of both types of ions. The pore with surface charge density of $-2$ e/nm$^2$ shows higher concentrations at negative voltages, but the depletion zone for positive 
voltages is less pronounced. In addition, the location of the depletion zone shifts towards the big opening of the pore. We postulate that the dependence of the rectification on the surface charge density is due to the same reason as the earlier reported maximum of the rectification degree on the bulk KCl concentration\cite{Cervera2006}. In the extreme case of zero concentration of the bulk electrolyte, a conical nanopore was predicted to be able to conduct only cations and as a result, currents for voltages of both polarities became equal to each other. In a pore with a given surface charge density, lowering the salt concentration beyond the maximum enriches a larger part of the pore with potassium pushing the location of the depletion zone towards the large opening. Increase of the surface charge density also causes an enhancement of the pore ionic selectivity for both voltage polarities, and shift of the position of the depletion zone. The depletion zone is also characterized by higher ionic concentrations compared to less charged pores, because the cations are sourced from the side with the large pore opening through a shorter resistive element.

Our results clearly indicate that a rectifying conically shaped nanopore cannot be approximated by a structure with a cation selective tip, as suggested by Momotenko et al\cite{Momotenko2011}. The relation between rectification and selectivity is very complex and arises from the voltage-dependent ionic selectivity and ionic concentrations in the pore.

\subsection{Influence of the pore length on the rectification behavior}\label{e:porelength}

\begin{figure}[h!]
\subfloat[\label{f:iv_len}]{\includegraphics[width=0.47\textwidth]{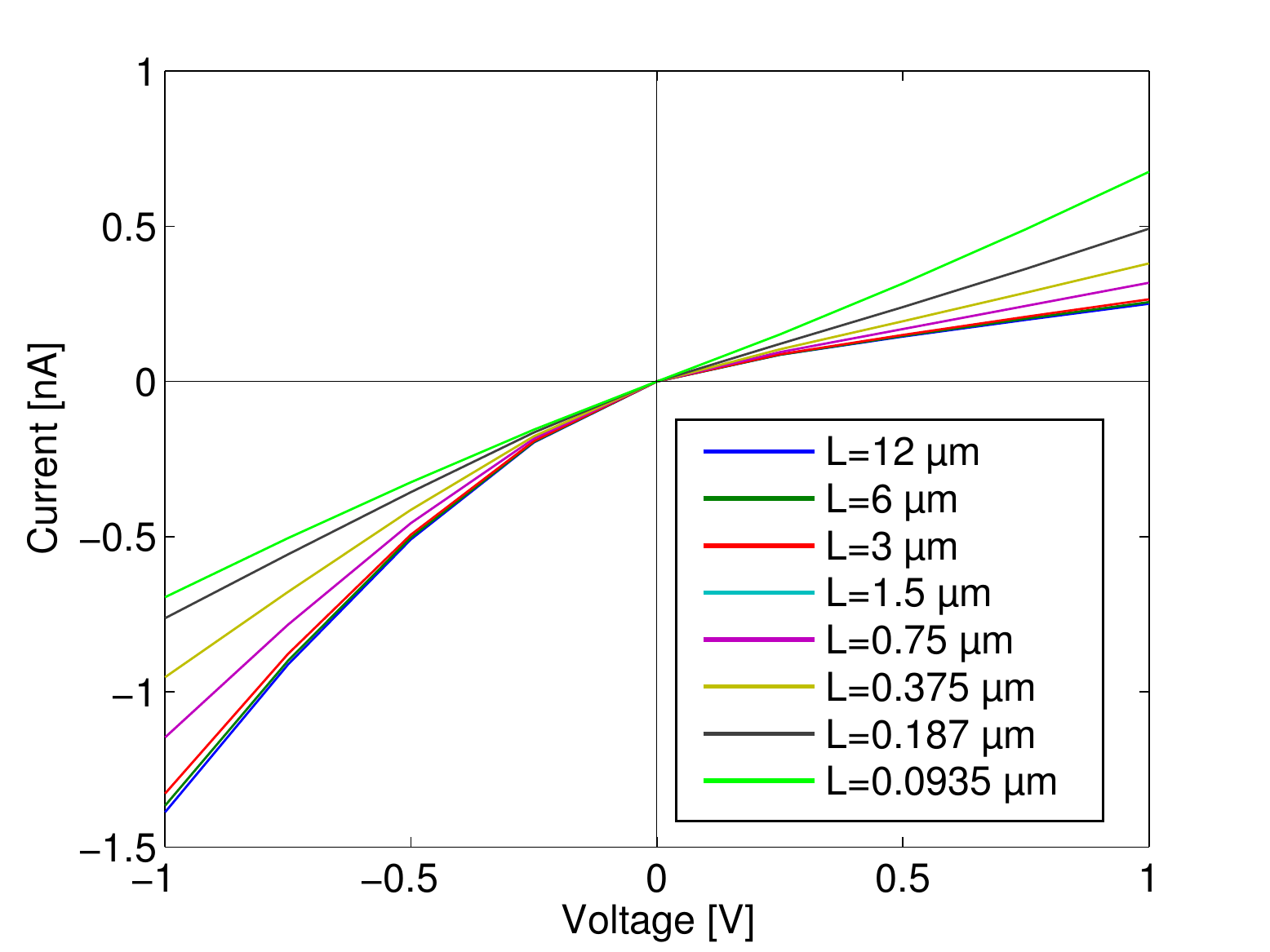}} % [I-V curves for different lengths. Aspect ratio is kept constant. 

\subfloat[\label{f:rect_len}]{\includegraphics[width=0.47\textwidth]{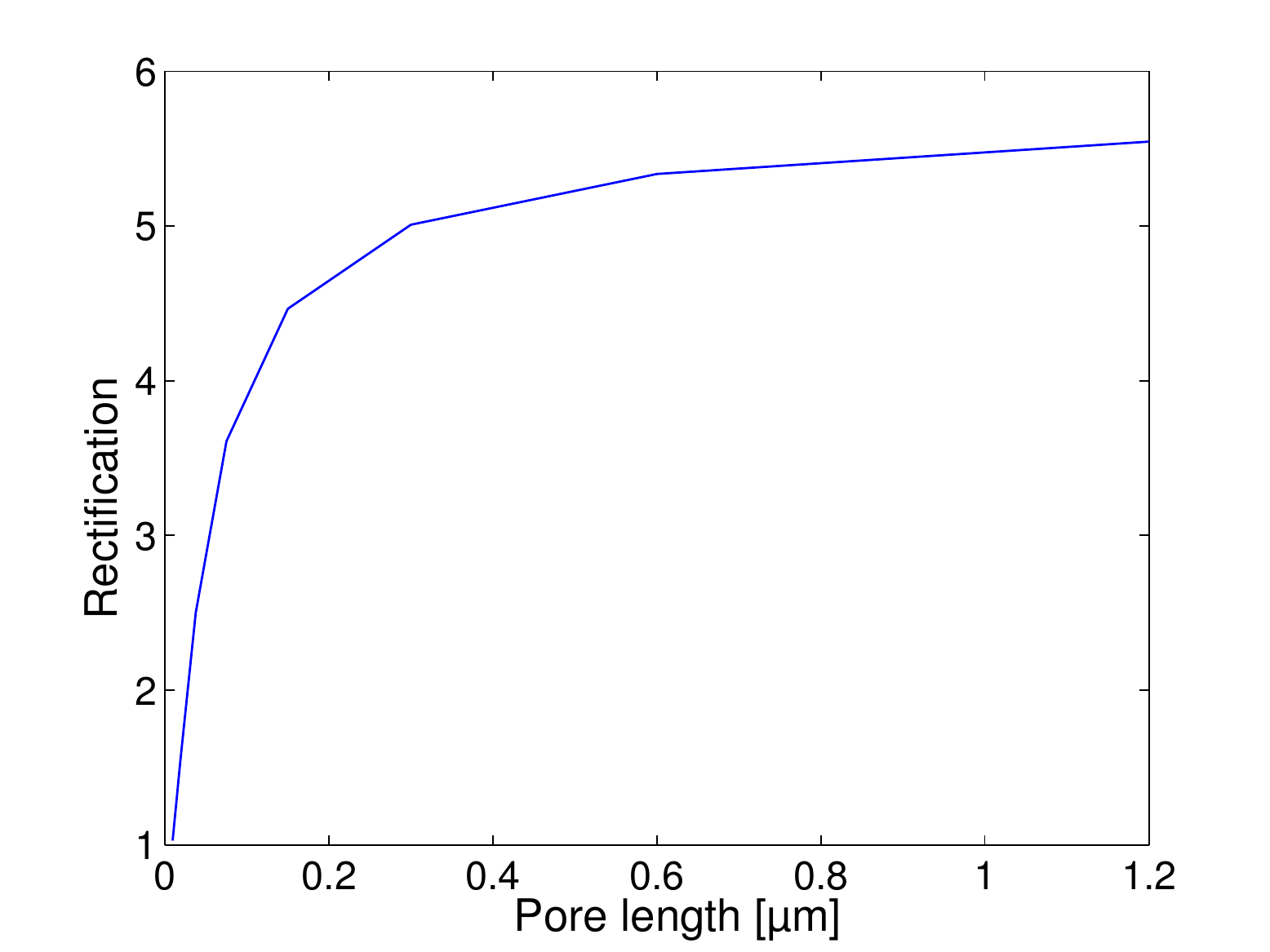}} % Rectification as a function of pore length 
% \caption{IV curves for different length and corresponding rectification. The small opening of the pore is kept at 6 nm and $\sigma = -1\text{e}/\text{nm}^2$.}
\caption{(a) Calculated current-voltage (I-V) curves for conical pores with a small opening of 6 nm, surface charge density of -1 e/nm2 and different lengths as indicated in the figure. (b) Rectification degrees for pores whose I-V curves are shown in (a) for currents at 1V.}
\label{f:iv_length}
\end{figure}

\begin{figure}[h!]
\subfloat[\label{f:densities_length_long}]{\includegraphics[width=0.51\textwidth]{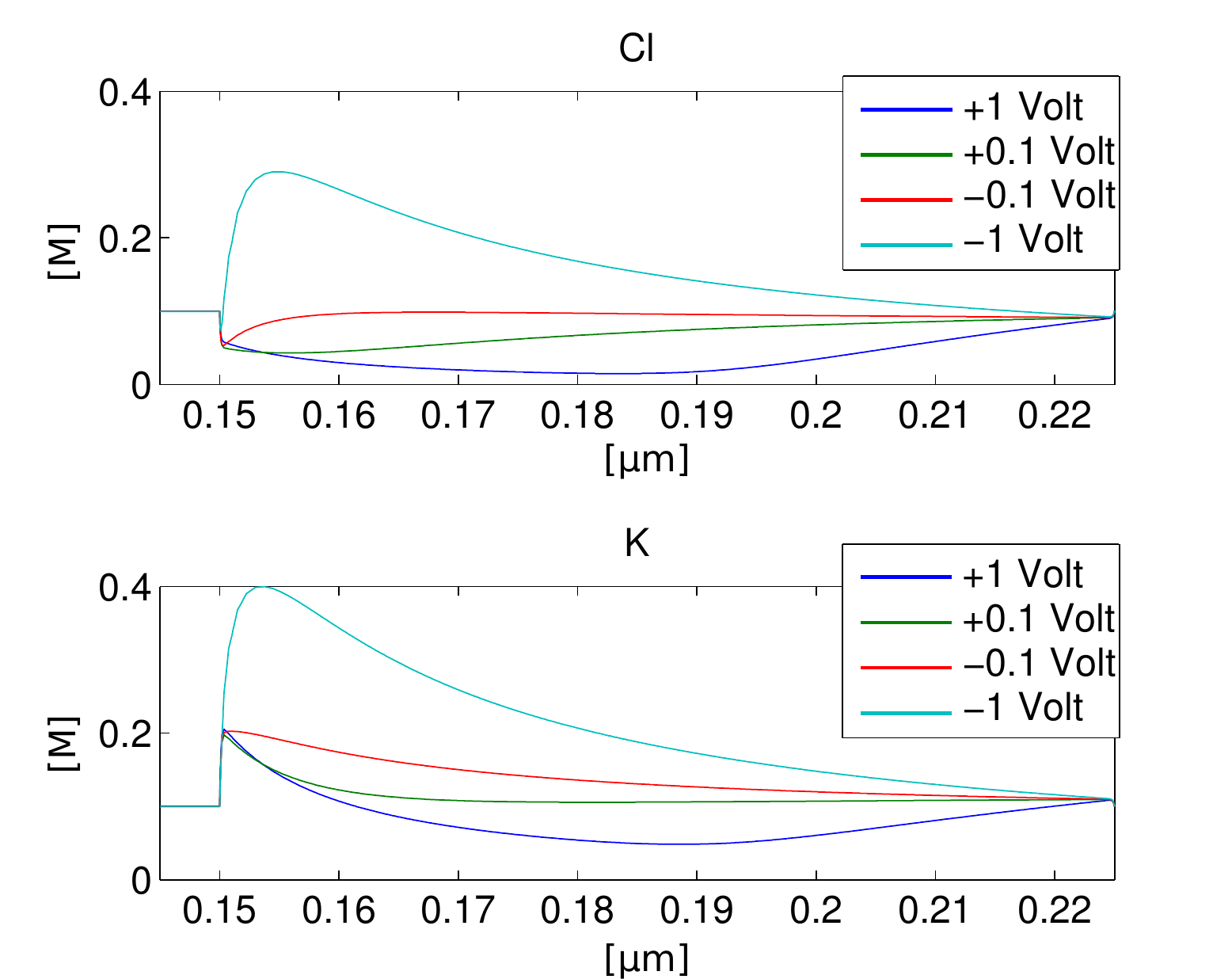}}

\subfloat[\label{f:densities_length_short}]{\includegraphics[width=0.51\textwidth]{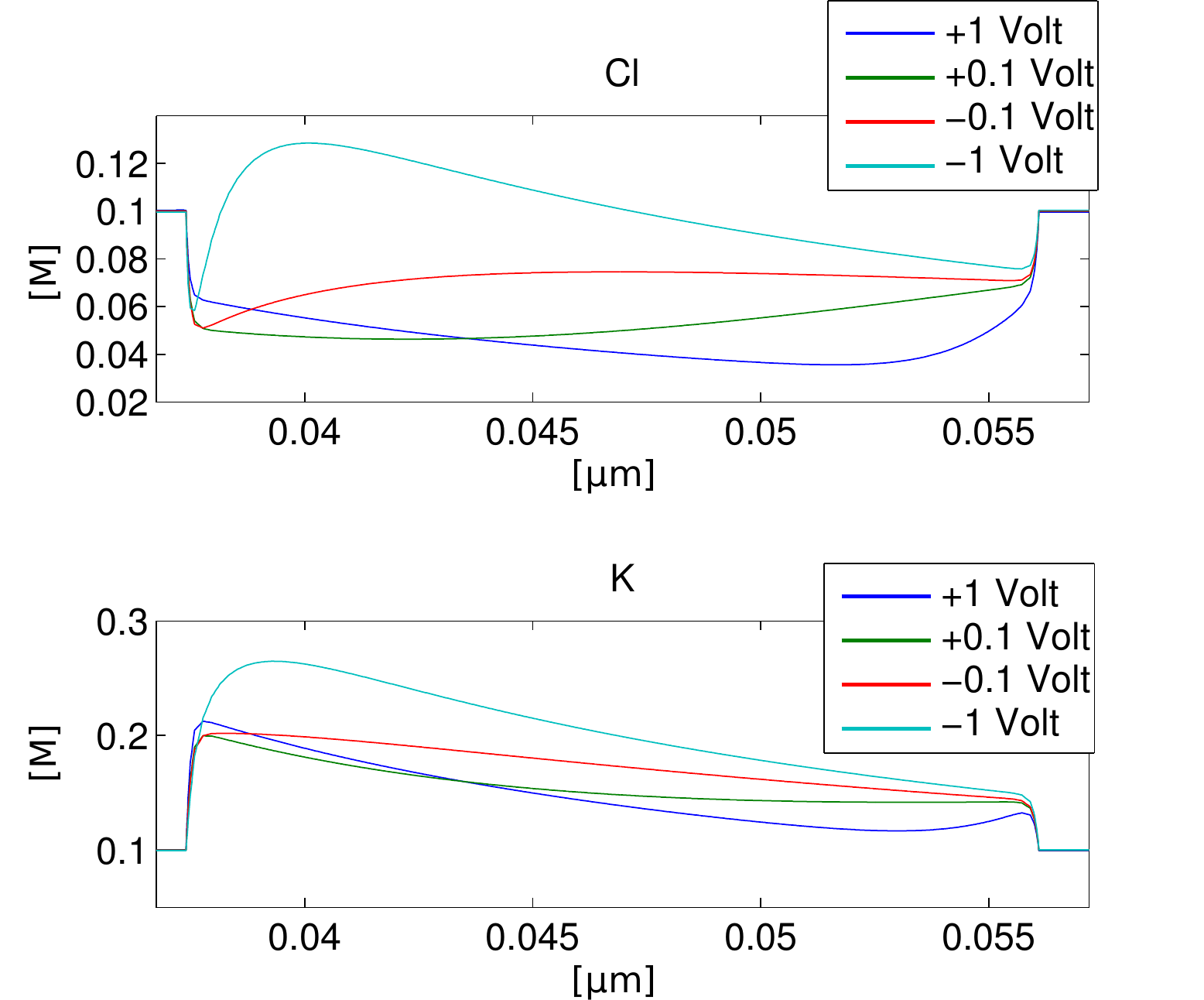}}
% \caption{Densities inside the pore region for a conical pore with length $L = 750\;\mathrm{nm}$ and $L = 187\;\mathrm{nm}$ and surface charge density $\sigma = -0.14\;e/nm^2$ for voltages -1 V, -0.1 V, 0.1 V and 1 V}
\caption{Ionic concentrations along the pore axis for conical pores with length of (a) $750$ nm, and (b) $187$ nm and surface charge density $\sigma = -0.14\;e/nm^2$ for different voltages as indicated in the figure. Bulk KCl concentration was $0.1$ M KCl.}
\label{f:densities_length}
\end{figure}

\begin{figure}[h!]
\subfloat[]{\includegraphics[width=0.47\textwidth]{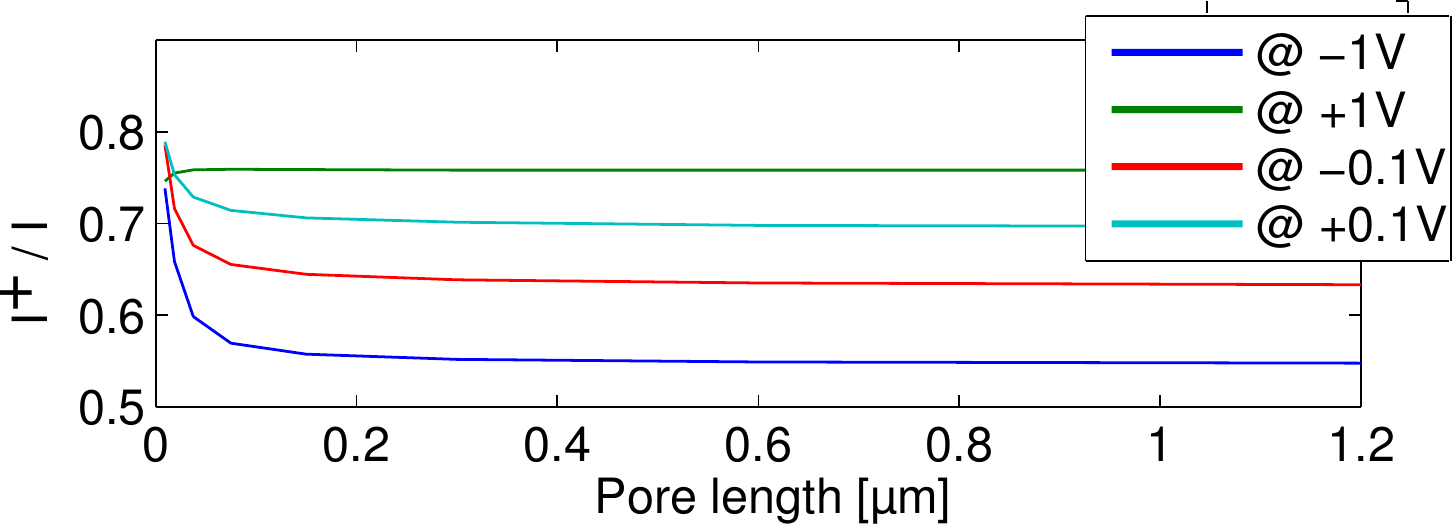}} % [I-V curves for different lengths. Aspect ratio is kept constant. 

\subfloat[]{\includegraphics[width=0.47\textwidth]{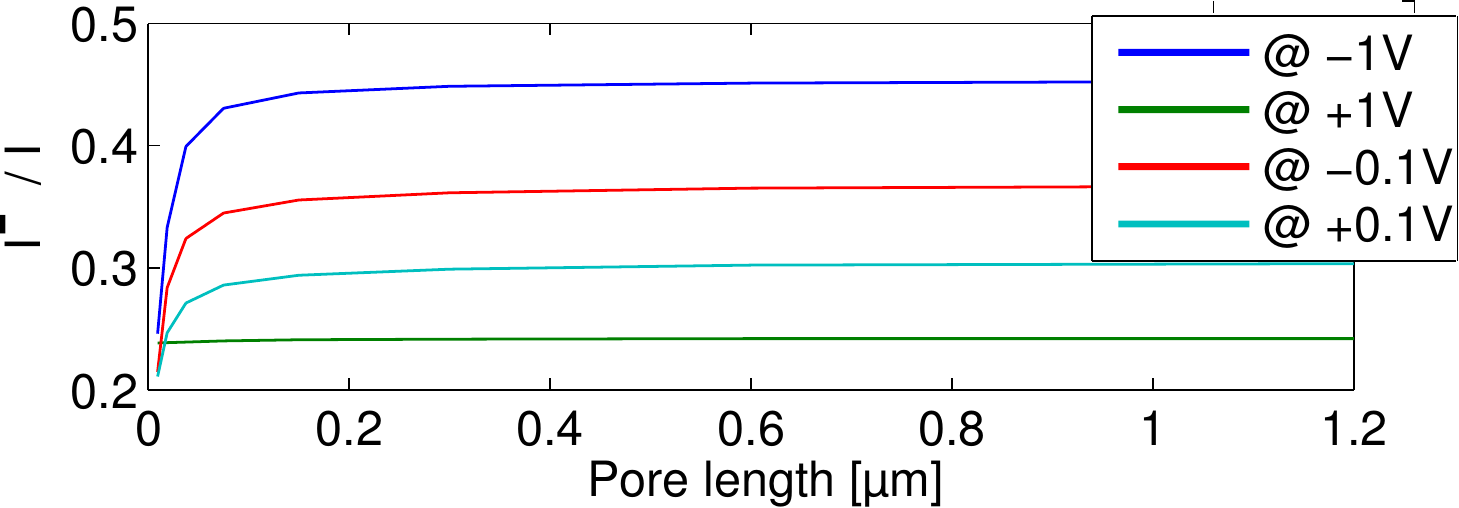}} % Rectification as a function of pore length 

\caption{(a) Ratio of the flux of potassium ions (denoted by $I^+$) and the total current (I), and (b) Ratio of the flux of chloride ions (denoted by $I^-$) and the total current (I), for pores of different lengths and different voltages. All simulations were done for a surface charge density of $-0.14$ e/nm$^2$ and $0.1$ M KCl.}\label{f:ratiocurrent}
\end{figure}

There has been recent interest in the preparation of asymmetric pores in membranes of various thicknesses, thus we investigated the dependence of the current-voltage curves of conically shaped pores on the pore length\cite{Powell2011,Vlassiouk2009b,Perry2012}. The corresponding rectification degrees were calculated based on currents recorded at $\pm $ $1$ V for $0.1$ M KCl. The small opening diameter of the pore was kept constant at $6$ nm, and the big opening was changed accordingly, but keeping the cone opening angle the same as the pore shown in Fig. \ref{f:6nm812nm}. Table \ref{t:lengths} lists the sets of pore lengths and opening radii used in the simulation. The rectification degree was found to decay with the decrease of the pore length, which as suggested by the simulated current-voltage curves stems from the rapid increase of the currents for positive voltages, and decrease of negative currents. (Fig. \ref{f:iv_length}(a) and \ref{f:iv_length}(b) ). Figs. \ref{f:densities_surfcharge_low} and \ref{f:densities_length_long} and \ref{f:densities_length_short} show that indeed by making the pore shorter, the depletion zone for $+1$ V is 
much less complete leading to the increase of positive currents. The geometrical resistance of the pore decreases as well. The counterintuitive decrease of negative currents can be understood by the voltage dependence of ionic concentrations in the pore. Shorter pores experience weaker enhancement of potassium and chloride concentrations compared to a long pore, and as a result, negative currents gradually decrease with the decrease of the pore length (Figs. \ref{f:densities_length}(a) and \ref{f:densities_length}(b)). With decreasing  pore length, the concentration of chloride ions diminishes more rapidly than the concentration of potassium ions. As a result, shorter pores are cation selective even at -1V (Fig. \ref{f:ratiocurrent}). \\

\begin{table}[h!]
\begin{center}
\caption{Different pore lengths and corresponding opening radii used for the numerical simulations.}
		\begin{tabular}{ccc}
		Pore Length [$\mu m$] & $r_{\mathrm{small}}$ [nm] & $r_{\mathrm{large}}$ [nm]\\\hline
		12 & 3 & 406\\
		6 & 3 & 204.5\\
		3 & 3 & 102.3\\
		1.5 & 3 & 51.1\\
		0.75 & 3 & 25.6\\
		0.375 & 3 & 12.8\\
		0.187 & 3 & 6.4\\
		0.0935 & 3 & 3.2\\
		
		\end{tabular}
		\label{t:lengths}
\end {center}
\end{table}

\begin{figure}[h!]
\subfloat[]{\includegraphics[width=0.5\textwidth]{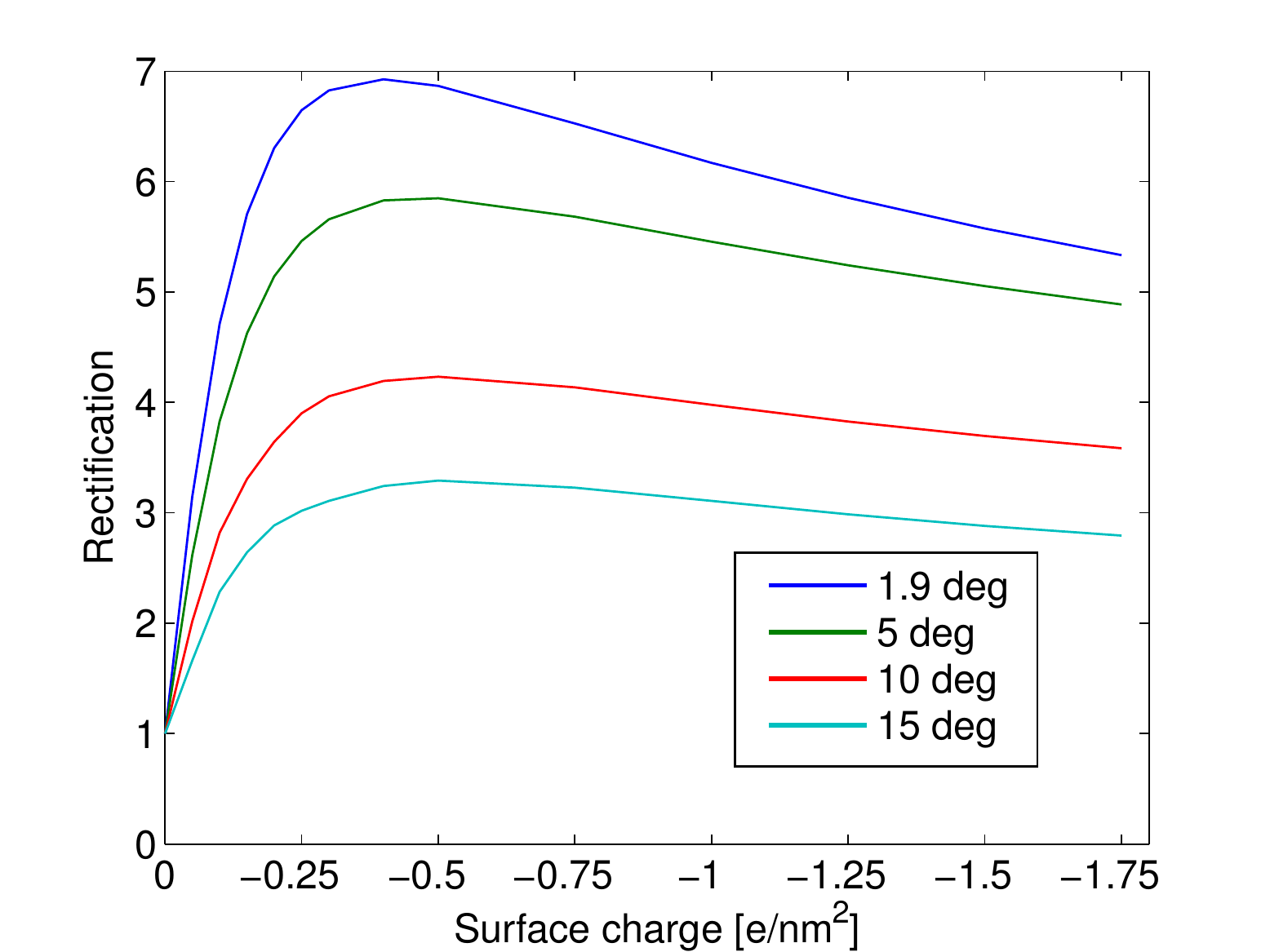}}

\subfloat[]{\includegraphics[width=0.5\textwidth]{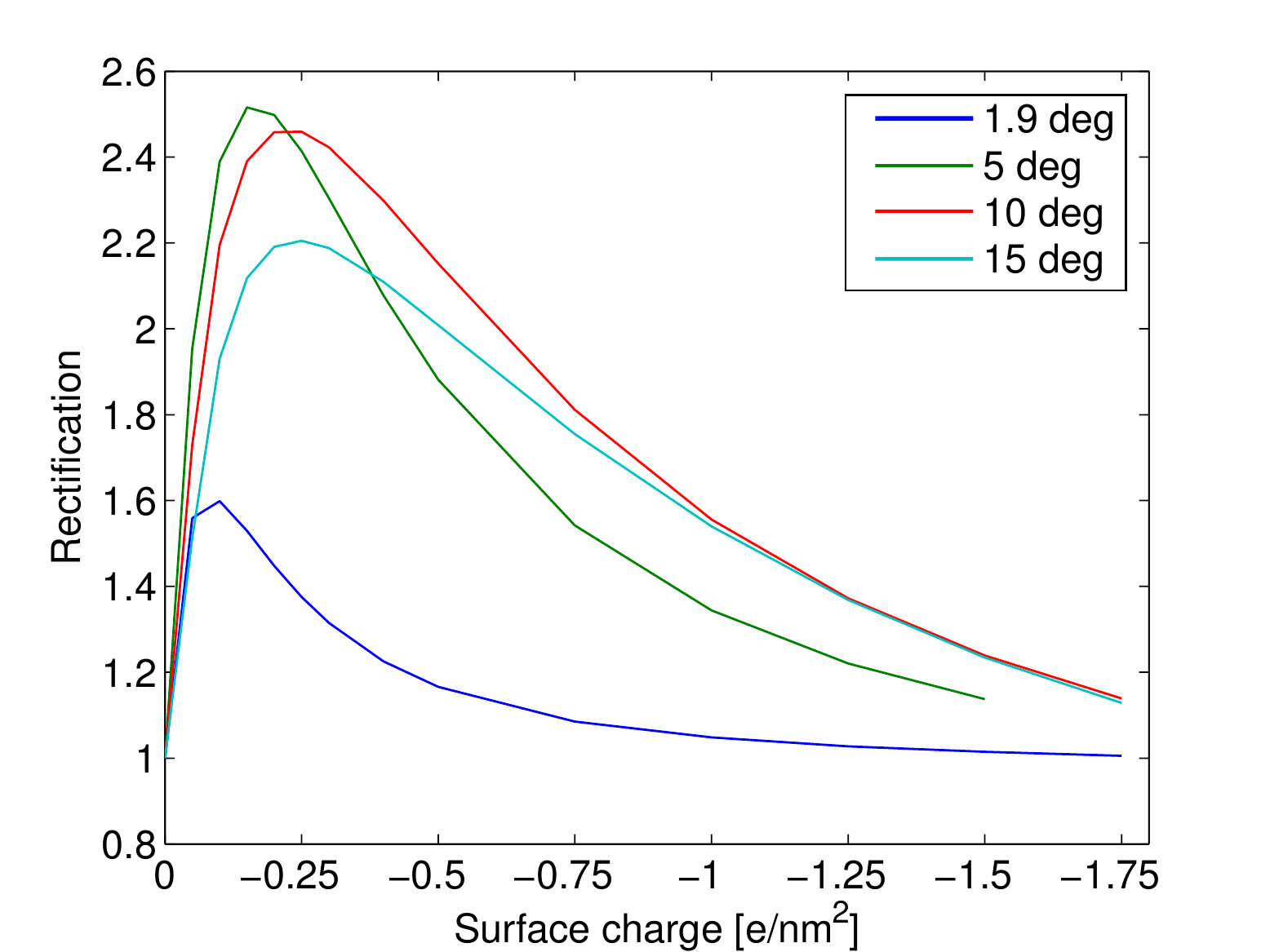}}
\caption{Rectification degree as a function of surface charge density and cone opening angle for pores with length of (a) $12$-$\mu$m, and (b) $187$ nm. The small opening diameter was kept $6$ nm in all simulations}\label{f:sel_surfcharge}
\end{figure}

The finding of an optimum surface charge, for which a maximum value of rectification is observed (Fig. \ref{f:sel_6nm812nm}(a)), initiated further computational experiments to understand whether the rectification properties of short pores can be improved by tuning the surface charge density of the pore walls and the cone opening angle. Fig. \ref{f:sel_surfcharge} confirms that these two parameters influence the rectification properties of $12$-$\mu$m and $187$ nm long nanopores. Interestingly, for the longer pore, the increase of the pore opening angle caused a decrease of the rectification properties. The influence of the opening angle on rectification of the $187$ nm pore is different. The lowest rectification has been predicted for the smallest opening angle, and the difference in rectification for pores with $5$, $10$, and $15$ degrees opening angle was rather small. The short pore also exhibited a more significant dependence of the rectification on the surface charge compared to the $12$-$\mu$m long 
structures (Fig. \ref{f:sel_6nm812nm} and Fig. \ref{f:sel_surfcharge}).\\

\begin{figure}[h!]
\subfloat[\label{f:rect_curvature_large}]{\includegraphics[width=0.5\textwidth]{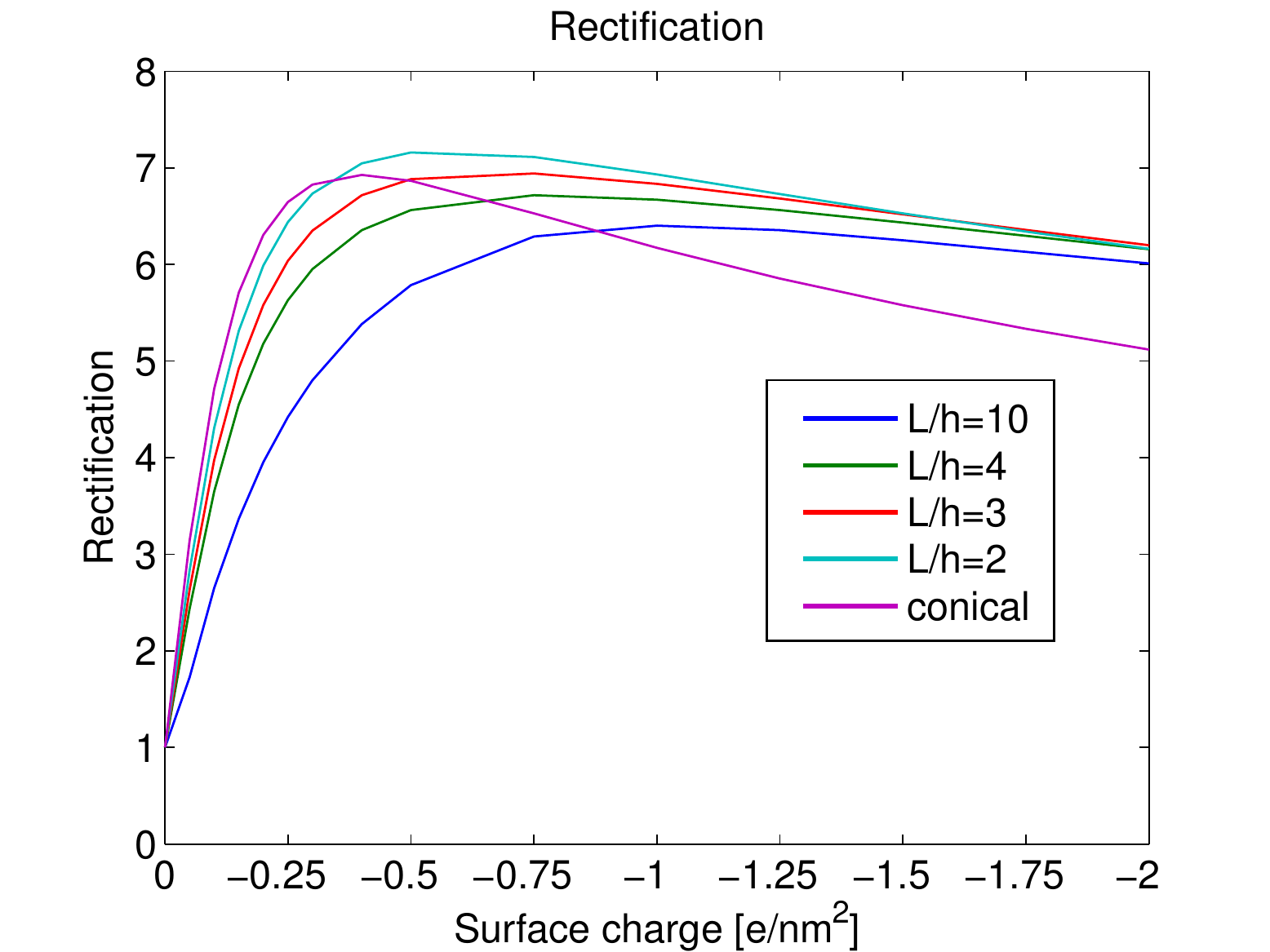}}

\subfloat[\label{f:rect_curvature_small}]{\includegraphics[width=0.5\textwidth]{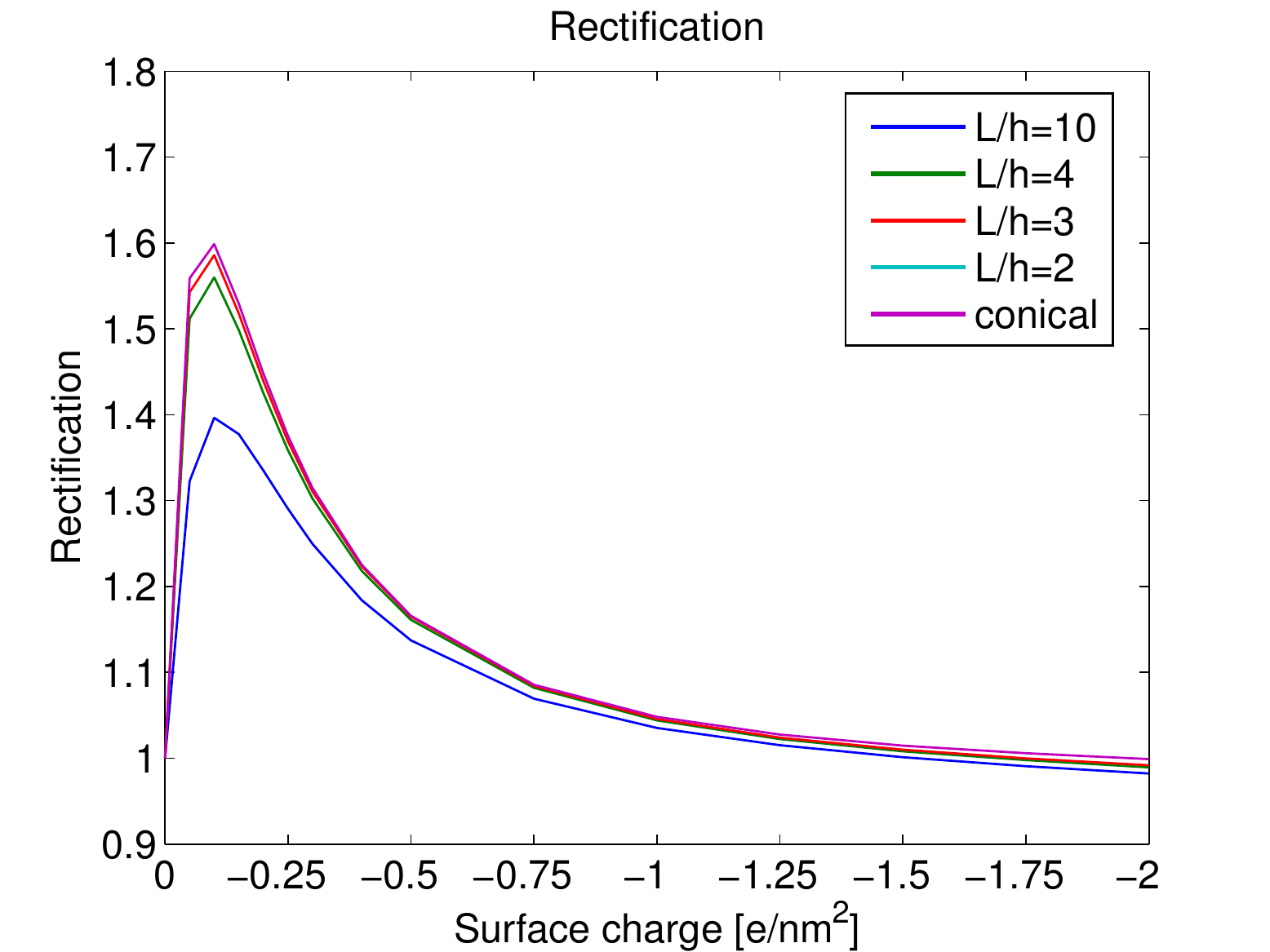}}
\caption{Rectification degrees for pores of length (a) $12$-$\mu$m, and (b) $187$ nm as a function of surface charge density and curviness defined by the parameter h (Fig. \ref{f:area}). Bulk KCl concentration was $0.1$ M KCl.}\label{f:rect_curvature}
\end{figure}

Another parameter, which was experimentally demonstrated to influence current-voltage curves of asymmetric pores is the curviness $h$ as shown in Fig. \ref{f:area}, cf. \cite{Ali2010,Apel2011}. Making a pore cigar-shaped was found to increase the rectification for long pores, at least for surface charges higher than $\approx -0.5$ e/nm$^2$, see Fig. \ref{f:rect_curvature_large}. Fig. \ref{f:rect_curvature_small} investigates whether a similar improvement can be observed in short pores. Changing the parameter $h$ of short conically shaped pores was found to have a smaller effect of the rectification compared to the effect of surface charge density. Increasing the value $L/h$ to four improved the rectification by an insignificant amount.

\section{Conclusion}
In this work we present detailed studies of transport properties of homogeneously charged conically shaped pores using the novel software $1$D solver MsSimPore\cite{MsSimPore}. We focus on how current rectification depends on the pore length, when keeping the pore opening diameter constant. Understanding the effect of the longitudinal dimension is important due to the ongoing miniaturization of ionic devices and efforts of making ionic circuits\cite{Han2009,Sannomiya2010,Tybrandt2012}. \\

The ability of a conically shaped pore to rectify the current was found to decrease with decreasing pore length. This observation is in accordance with experimental reports showing current-voltage curves of $12$ and $24$ $\mu$m long pores\cite{Apel2011}. When the pore length reached values below 1 $\mu$m, the pores showed an ohmic current-voltage curve. Short pores can however regain rectification if the surface charge density of the pore walls is appropriately tuned. We found out that the surface charge density, for which the maximum rectification is observed,  depends on the pore opening angle.\\

The modeling and simulation of ion currents and ionic concentration profiles in conically shaped pores also allowed us to provide understanding on the relationship between the pores’ rectification and ion selectivity, thus the ability to transport only counterions. In contrast to earlier reports, we consider voltage-dependent ionic selectivity, which we find to be crucial for the ion current rectification. We also show that the maximum ionic selectivity does not guarantee maximum rectification properties. The simulations reveal that in order to capture the full picture of physical phenomena underlying ion current rectification in conically shaped pores, the whole pore length has to be considered. A proposed reduced model of a conical pore containing a perfectly ion selective plug, although predicting rectification cannot describe the voltage-dependence of ionic selectivity and ionic concentrations in the pore\cite{Momotenko2011}.\\
  
All presented results have been performed using the software package MsSimPore, which is based on the Poisson-Nernst-Planck equations. The software was developed for conical and cigar shaped nanopores, which are characterized by high aspect ratios and surface charge densities. MsSimPore allows for reliable, stable and efficient simulations of concentration and voltage profiles as well as IV and rectification curves. The software can therefore run on a regular PC and is applicable to structures whose simulation would normally require the use of powerful workstations. The presented model, although one-dimensional, includes two electrolyte reservoirs in contact with the pore openings,
which allows for a more accurate description of the ionic currents than previously presented PNP models.\\

\noindent Further developments of MsSimPore will focus on identification problems, where structural parameters of nanopores are determined from transport characteristics. This is very important since non-destructive ways of pore imaging are limited. The geometry of polymer nanopores is usually studied by preparing their metal-replica \cite{Martin1994,Scopece2006,Toimil2012}. This approach can not reveal the structure of the narrow tip and works best for pores that are at least several tens of nanometers in diameter. Current-voltage curves do not only depend on the pore geometry but also on the electrolyte modulated surface characteristics of the pore walls.  Therefore the identification of the proper surface charge is necessary to identify the pore shape correctly. First results on inverse problems for ion channels, and a comprehensive overview and lookout on problems and methods for identification problems in synthetic and biological pores was given by Burger et al. \cite{Burger2007,Burger2011,Engl2012}. We also plan to introduce the finite size of ions, which will be especially important for pores characterized by extremely high surface charges densities and possible resulting crowding of ions. Considering size of ions will also be crucial for the description of even shorter pores whose length approaches the pore opening diameter, as it is the case of biological channels\cite{Boda2007,Boda2008}. To this end, we also refer to a non-linear variant of the PNP-Equations\cite{Burger2010}, which have already been applied in the context of ion channels\cite{Burger2012}.

This future direction of research is a promising application of collaborative research between experimental physicists and applied mathematicians, which will lead to new insights into the properties and behavior of nanopores.\\

%Another direction of research are the generalizations of the PNP equations have been proposed in the literature recently. In the classical d%escription ions are treated as point charges, newer approaches incorporate the physical size of ions in the modeling. These size effects are% important in very narrow channels and lead to nonlinear mobilities and diffusivities in the PNP equations\cite{Burger2012,Cervera2010,Zhou2%011}. Gillespie et al.  finite size effects are included via an additional potential in the transport equation\cite{Gillespie2002}. Another %generalization of the PNP equations in 2D, which includes chemical reactions to describe the formation of nanoprecipitates and resulting ion% current oscillations in conical nanopores was presented by Wolfram et al.\cite{Wolfram2010}.


\footnotesize{
\bibliography{pccp} %your .bib file
\bibliographystyle{rsc} %the RSC's .bst file
}

\end{document}